\documentstyle[preprint,aps,tighten]{revtex}
\begin{document}
\draft
\title{Screening enhancement factors for laboratory CNO and rp astrophysical
reactions.}
\author{Theodore E. Liolios \thanks{%
www.liolios.info}}
\address{Hellenic Naval Academy of Hydra\\
School of Deck Officers, Department of Science\\
Hydra Island 18040, Greece}
\maketitle

\begin{abstract}
Cross sections of laboratory CNO and rp astrophysical reactions are enhanced
due to the presence of the multi-electron cloud that surrounds the target
nuclei. As a result the relevant astrophysical factors are overestimated
unless corrected appropriately. This study gives both an estimate of the
error committed if screening effects are not taken into account and a rough
profile of the laboratory energy thresholds at which the screening effect
appears. The results indicate that, for most practical purposes, screening
corrections to past relevant experiments can be disregarded. Regarding
future experiments, however, screening corrections to the CNO reactions will
certainly be of importance as they are closely related to the solar neutrino
fluxes and the rp process. Moreover, according to the present results,
screening effects will have to be taken into account particularly by the
current and future LUNA experiments, where screened astrophysical factors
will be enhanced to a significant degree.
\end{abstract}


\section{Introduction}

Proton-induced nuclear reactions in stellar plasmas play a crucial role in
advanced stellar nucleosynthesis. For example the r(apid) p(roton-capture)
process\cite{walwoo,wienova} is the dominant reaction sequence in
high-temperature hydrogen burning. Such processes occur when hydrogen fuel
is ignited under highly degenerate conditions in explosive events on the
surface of white dwarfs, neutron stars and ordinary supermassive
stars.\thinspace The proton-rich nuclei which participate in the rp process
are currently under study in various laboratories around the world while
much interest has attracted the prospect of studying relevant proton capture
reactions for radioactive nuclei in radioactive ion-beam facilities. The
beam energy in such reactions is sometimes so low (astrophysical energy)
that the proton impinging on the multielectron target will actually
experience an acceleration due to the electron cloud of the target.

A recent study\cite{lioliossef} derived an analytic screening enhancement
factor for binary multielectron reactions (between atoms $%
_{Z_{2}}^{A_{1}}X\,\,$and $_{Z_{2}}^{A_{1}}X\,$\thinspace at a relative
energy $E\,$), which is actually a corrective factor $f\left(
Z_{1},Z_{2},A_{1},A_{2},E\right) \,\,$defined as

\begin{equation}
f\left( Z_{1},Z_{2},A_{1},A_{2},E\right) \,=S_{\exp }\left( E\right)
/S\left( E\right)
\end{equation}
where $S_{\exp }\left( E\right) \,$is the experimental value of the
astrophysical factor while $S\left( E\right) \,$is the corresponding one for
bare nuclei (disregarding the electron cloud).

Thus experimentalists should actually plot (or tabulate) the values $S_{\exp
}\left( E\right) /$ $f$ \thinspace instead of the value $S_{\exp }\left(
E\right) .\,$However, many such astrophysical factors which have not been
corrected for screening have been used directly in the thermonuclear
reaction rates\cite{angulo}.

On the other hand, as shown recently\cite{liolioss0} the effort to lower the
energy beam close to zero (so that the value $S\left( 0\right) \,$is more
accurate) is pointless. The thermonuclear reaction rate is more accurate if
the value $S\left( E_{0}\right) $ is used instead of $S\left( 0\right) $%
\thinspace \thinspace and that is where the experimentalist should focus.
Nevertheless, if the uncorrected value of $S\left( E_{0}\right) $ is used
then the relevant reaction rate is of course overestimated by a factor of $%
f\left( Z_{1},Z_{2},A_{1},A_{2},E_{0}\right) \,\,$.

In Ref. \cite{lioliossef} the new corrective method was applied to
non-resonant reactions of the CNO cycle in typical solar conditions. In the
present study we extend that method to more advanced proton-induced
reactions (e.g. of the rp process) which, of course, are not expected to be
non-resonant ones; on the contrary some of them exhibit very dense patterns
of resonances at relatively high energies so that statistical methods such
as the Hauser-Feshbach one are needed in order to calculate the respective
reaction rate. However, when the compound nucleus exhibits low-level
densities then the statistical model breaks down\cite{wormer}. Therefore,
especially for proton-rich nuclei near the proton drip line with small
Q-values of proton capture reactions, the Maxwellian averaged reaction rate $%
N_{A}<\sigma u>$ \thinspace is determined by single isolated (narrow)
resonances

\begin{equation}
N_{A}<\sigma u>_{r}=N_{A}\left( \frac{2\pi }{\mu k}\right) ^{3/2}\hbar
^{2}\left( \omega \gamma \right) _{r}T^{-3/2}\exp \left( -\frac{E_{r}}{kT}%
\right)  \label{resterm}
\end{equation}
and their non-resonant (tail) contributions

\begin{equation}
N_{A}<\sigma u>_{tail}=N_{A}\left( \frac{2}{\mu }\right) ^{1/2}f_{pl}\frac{%
\Delta E_{0}}{\left( kT\right) ^{3/2}}S_{eff}\exp \left( -\frac{3E_{0}}{kT}%
\right)  \label{tailterm}
\end{equation}
so that $N_{A}<\sigma u>$ \thinspace =$N_{A}<\sigma u>_{r}+N_{A}<\sigma
u>_{tail},$ where we have used the familiar notation of Ref. \cite{fowler67}
(i.e. $N_{A}$ is the Avogadro number, $\hbar $ is Planck's constant, $\left(
\omega \gamma \right) _{r}\,$is the usual quantity appearing in the
Breit-Wigner single-level formula, $k$ is Boltzmann's constant, $\mu $ is
the reduced mass, $E_{0}\,$is the most effective energy of interaction, $%
\Delta E_{0}$ the relevant energy window, $S_{eff}$ the effective
astrophysical factor, and $T$ is the temperature). It should be emphasized
that $f_{pl}$ is the screening enhancement factor due to plasma effects
which of course is calculated via entirely different models (see for example
Ref.\cite{liolios1} and references therein). The interplay between plasma
and laboratory screening enhancement factors has been thoroughly discussed
in Ref. \cite{liolioss0}

Note that despite the presence of resonances the concept of the most
effective energy of interaction $E_{0}$ still serves its purpose which is to
point to the energy region where a resonance will have dramatic effects on
the reaction rate.

If the quantity $S_{eff}=S\left( 0\right) $ ($S_{eff}=S\left( E_{0}\right)
\, $) $\,$is to be accurately determined then the beam energy should be as
low (as close to $E_{0})\,$as possible. In practice, however, it is not
feasible to lower the energy so much which means that either
experimentalists obtained $S_{eff}$ by extrapolating from higher energy
values (definitely higher than $E_{0})$ or they relied on theoretical models
(e.g.\cite{descouv}). In the first case, which is the most usual, a
substantial error is committed sometimes. The higher the energy of the last
measurement used in the extrapolation the further the experiment from the
astrophysically important region and, of course, the larger the error. Note
that by ''shooting'' from that far not only can we commit a statistical
error in the fitting procedure but, more importantly, one cannot possibly
know if there is a resonance close to $E_{0}\,$(above or below). If such
resonances exist and pass undetected then the implications to critical
astrophysical models can be enormous. As an example we refer to the $%
^{3}He\left( ^{3}He,2p\right) ^{4}He\,$\thinspace reaction\cite
{lunasecond,lioliosluna} and the relevant neutrino fluxes which, until
recently, suffered from such large uncertainties.

The need for an accurate $S_{eff}$ appearing in the non-resonant component
of the reaction rate indicates that it is urgent that we should provide
experimentalists with a clear profile of all the laboratory energies beyond
which the relevant astrophysical factor must be corrected before used in
stellar evolution codes. We will assume that the target is in a neutral
atomic state as well as that the energy of proton projectiles is roughly the
relative energy of the collision.

The former assumption needs some justification. Admittedly considering the
target to be in an atomic state is an undesirable simplification which has
been used by other similar studies as well\cite{bracci,braccilet}. Note
that, some times, the projectile itself is in a molecular state (e.g. $D_{2}$%
) and then the whole process becomes even more complicated. Actually, for a
full dynamical treatment of electronic degrees of freedom, the molecular
few-body problem, which takes into account explicitly the electrons and
nuclei of the system, should be properly solved. This is a very difficult
task and the only relevant study available\cite{shoppa} so far is referring
to hydrogenic molecules. The results indicate that the phenomenon is
extremely complex where, for example, even the orientation of the target
molecule plays a non-negligible role. Unfortunately, taking into account the
molecular nature of the targets involved in CNO and rp experiments is beyond
the ambitions of the present study, whose only objectives are: a) to
approximately define the energy region where screening effects are important
and, b) to provide an estimate of the screening enhancement. However, it is
very encouraging that Ref. \cite{shoppa} came to the conclusion that taking
into account the molecular nature of their target increases the relevant
screening energy. Thus, at first sight, considering the target to be in an
atomic state seems to be a conservative approach which adopts the smallest
possible screening energy. This is not the case, according to other studies%
\cite{langrolfs,engstler}, which subtract the molecular binding energy and
the take-away kinetic energy of the spectator nuclei from the total (atomic)
screening energy. In fact the situation is more complicated since the
''solid state'' effects are not fully understood nowdays and are still under
investigation. In view of the inability of all available theoretical models
to account for the large experimental screening energies the present study ,
by fulfilling its objectives (a,b), provides a good point of reference to
all future efforts to solve the screening puzzle.

Under the previously defined assumptions the screening enhancement factor of
laboratory proton-induced reactions is confined between the sudden limit
(SL)\ and the adiabatic limit (AL) so that $f_{SL}<f<f_{AL},\,$where the two
limits for neutral targets can be derived if we modify appropriately the
respective formulas of Ref.\cite{lioliossef}:

Sudden Limit:

\begin{equation}
f_{SL}\left( E\right) \simeq \exp \left[ \frac{0.765Z_{1}^{7/3}A^{1/2}}{%
E_{\left( keV\right) }^{3/2}}\right]  \label{slsef}
\end{equation}

Adiabatic Limit:

\begin{equation}
f_{AL}\left( E\right) \simeq \exp \left[ \frac{0.3176A^{1/2}}{E_{\left(
keV\right) }^{3/2}}Z_{1}\left[ \left( Z_{1}+1\right)
^{7/3}-Z_{1}^{7/3}\right] \right]  \label{alsef}
\end{equation}
where$\,A$ is the reduced mass number $A=A_{1}A_{2}\left( A_{1}+A_{2}\right)
^{-1}.$

These formulas have been derived in the framework of the Thomas-Fermi model,
taking into account exchange, ionization, thermal, and relativistic effects
of the atomic cloud.

It is easy to show that when $Z_{1}>8$ , as is usually the case for advanced
proton-induced astrophysical reactions, the gap between the above two limits
narrows considerably thus providing an excellent constraint for the
respective screening enhancement factor. If we also take into account that
for such massive targets $A\simeq 1$ then the SEF can be written

\begin{equation}
f\left( E\right) \simeq \exp \left( 0.765Z_{1}^{7/3}E_{\left( keV\right)
}^{-3/2}\right)  \label{psef}
\end{equation}
and of course describes the relevant screening enhancing effect in the lab
in a very accurate way.

In order to accurate measure $S_{eff}$ ($S\left( E_{0}\right) \,$\thinspace
or $S\left( 0\right) $) nuclear astrophysics experiments have to be carried
out (at or lower than) the most effective energy of interaction which for
the reactions considered here takes the simple approximate form$:$

\begin{equation}
E_{0}\simeq 1.22(Z_{1}T_{6})^{2/3}keV  \label{moefrp}
\end{equation}
In figure 1 we plot the most effective energy of interaction given by Eq.$%
\left( \ref{moefrp}\right) \,$with respect to plasma temperature for various
proton-induced thermonuclear reaction. This is the energy at (or below)
which the experiment has to be carried out in order to accurately determine
the value of $S\left( E_{0}\right) \,$\thinspace (or $S\left( 0\right) $)$.$
In figure 2 we plot the laboratory screening enhancement factor with respect
to the relative energy of interaction (center of mass) for various
proton-induced thermonuclear reactions with $Z_{1}=6,10,15,20,25,30,35,40\,$%
\thinspace . We have focused on the atomic effects threshold for each
reaction so that we can provide the experimentalist with the limit beyond
which the present screening corrections become important.

Now let us investigate the importance of the corrections proposed discussing
some particular experiments which although have already been contaminated by
the screening effect their results have not been corrected.

1) The $S_{eff}$ value of the reaction $^{14}N\left( p,\gamma \right) ^{15}O$
\thinspace adopted in Ref.\cite{angulo} is the one obtained in Ref. \cite
{schroed}\thinspace , which incorporated data\cite{lamb} contaminated by the
screening effect. At $T_{9}<1$ (i.e. $E_{0}<446\,keV)\,\,\,$the reaction is
dominated by the $1/2^{+}$ resonance at $E_{r}=259.4\pm 0.4\,keV\,\,$which
has a resonance strength $\left( \omega \gamma \right) _{r}=0.014\,eV$. If
for simplicity we ignore the effects of subthreshold resonances then the
relative contribution of the resonant term (Eq. $\left( \ref{resterm}\right)
)\,\,$and the tail one (Eq. $\left( \ref{tailterm}\right) $will be

\begin{equation}
N_{A}<\sigma u>_{r}=2389.5T_{9}^{-3/2}\exp \left( -3T_{9}^{-1}\right)
\,cm^{3}mol^{-1}s^{-1}
\end{equation}

\begin{equation}
N_{A}<\sigma u>_{tail}=1.\,5329\times 10^{10}S_{eff}T_{9}^{-2/3}\exp \left(
-15.193T_{9}^{-1/3}\right) \frac{cm^{3}mol^{-1}s^{-1}}{MeV\cdot b\,}
\end{equation}
Ref.\cite{angulo} adopts the value $S_{eff}=S\left( 0\right) =\left( 3.2\pm
0.8\,\right) 10^{-3}\,MeV\cdot b\,\,$admitting a systematic error of 25\% in
the calculation of $\,N_{A}<\sigma u>_{tail}.\,$However, the screening error
committed in Ref. \cite{lamb} has been neglected thus overestimating $%
S\left( 0\right) $ by a certain degree. The implications of neglecting the
screening enhancement can be very significant, especially if one is
interested in solar reaction rates. In fact the value $S\left( 0\right) $ in
question relied on measurements taken$\cite{lamb}$ at energies as low as $%
E_{cm}=93\,keV.\,\,$The astrophysical factor measurements at such energies
according to our SEF are overestimated by a factor $f_{^{14}N+p}\left(
93keV\right) =1.083.\,\,$To assess the significance of such errors let us
naively assume that the 8.3\% error committed at $E_{cm}=93$ is the actual
error in the evaluation of $S\left( 0\right) $ ( note that the measurements
of Ref. \cite{lamb}\thinspace suffer more from the fact the its energy is
far from the most effective energy of interaction than from lack of
screening corrections). This assumption is particularly valid for future
experiments (with $E_{cm}<93\,keV)\,\,$where the screening error will
naturally be much larger than $8\%\,(\,$thus rendering our assumption a
conservative one). As regards the solar neutrino problem an $+8\%$ error in
the value of $S_{^{14}N+p}\left( 0\right) $ leads to a similar
overestimation of \thinspace the theoretical values of the neutrino fluxes
generated by the solar reactions $^{13}N\left( e^{+}\nu _{e}\right) ^{13}C$
and $^{15}O\left( e^{+},\nu _{e}\right) ^{15}N,\,$thus accentuating
unnecessarily the discrepancy between theory and experiment. The
significance of such corrections as the ones we propose here was underlined
in a recent work\cite{bahcall} where it was pointed out that the solar model
predictions for the CNO fluxes are not precise because the CNO fusion
reactions are not as well studied as pp reactions. In the same work we learn
that the current $1\sigma $ error in the standard model CNO neutrino fluxes
is $17\%-25\%$ that is much larger than the respective error in the $pp$
reactions. Since an accurate measurement of the CNO neutrino fluxes would
constitute a stringent test of stellar evolution theory\cite{bahcall} the
corrections proposed in the present paper are particularly relevant.

Moreover, an $+8\%$ error in the value of $S_{^{14}N+p}\left( 0\right) $
causes an underestimation of the mean lifetime of $^{14}N$ in the hot CNO\
cycle (the generator of the rp process) which translates into an
overestimation of the production rate of $^{15}O$ which initiates the rp
process via the $^{15}O\left( \alpha ,\gamma \right) ^{19}Ne\,\,\,$reaction.

The need for a more accurate astrophysical factor for the reaction in
question is now obvious. Any future experiment aimed at increasing the
accuracy of $S\left( 0\right) \,$by lowering the beam energy will certainly
commit errors larger than 10\% rendering the present SEF indispensable.
According to the most recent NuPECC report\cite{nupecc} the new 400 kV
accelerator of the LUNA collaboration at the Gran Sasso is currently being
used for the measurement of the cross section of the reaction $^{14}N\left(
p,\gamma \right) ^{15}O\,$ at 70 keV. According to the present paper the
screening corrections at such temperatures will be of the order of 13\%.

2) The reaction $^{19}F\left( p,\alpha \gamma \right) ^{16}O\,\,$has a
number of resonances in the astrophysically important energy region, while
recently a new one was discovered$\left( \cite{haris}\right) $ at $%
E_{r}=237\,keV.$ In the relevant experiment the beam energy had to be
lowered below the atomic effects threshold so that some of its astrophysical
factor measurements were enhanced and need to be corrected. Thus, the low
energy values of $S_{\exp }\left( E\right) $ tabulated in Ref. $\left( \cite
{haris}\right) $should be replaced by $S_{\exp }\left( E\right) /f\left(
E\right) $ so that:

$
\begin{tabular}{llllllll}
$E_{cm}\,\left( keV\right) $ & 188.8 & 198.3 & 203.1 & 231.0 & 235.8 & 240.5
& 250.1 \\ 
$S_{\exp }\left( E\right) \,\left( MeV\cdot b\right) \cite{haris}$ & 5$\pm 4$
& 12$\pm 3$ & 26$\pm 4$ & 20$\pm 6$ & 22$\pm 5$ & 29$\pm 5$ & 33$\pm 5$ \\ 
$S\left( E\right) _{\exp }/f\left( E\right) \,\left( MeV\cdot b\right) $ & 
4.7$\pm 4$ & 11.4$\pm 3$ & 24.8$\pm 4$ & 19.2$\pm 6$ & 21.2$\pm 5$ & 28.0$%
\pm 5$ & 31.9$\pm 5$%
\end{tabular}
$

The largest correction occurs at the lowest energy $\left(
E_{cm}=188.8\,keV\right) \,$and it is of order\thinspace 5\%. Thus, so far,
no significant screening error has been committed in the experiment in
question. However as admitted in Ref. $\cite{haris}\,$\thinspace the
S-factor is still uncertain due to lack of experimental information at lower
energies. This uncertainty will naturally lead to experiments where the
energy should be much lower$.\,\,$At such low energies the experimental
measurements will be considerably enhanced and of course the present
screening enhancement factors may be important to the accuracy of the
experiment.

3) The reaction $^{86}Sr\left( p,\gamma \right) ^{87}Y\,$\thinspace , which
is important both to the p and the rp process, was recently investigated\cite
{harsr} in the astrophysically relevant energy range. In that experiment the
beam energy was lowered below the atomic effects threshold causing an
enhancement of the respective astrophysical factor. In fact the $S\left(
E\right) $ \thinspace value measured at the lowest energy of the experiment $%
S\left( 1477\,keV\right) $ has been overestimated by $6.7\%$\thinspace
\thinspace while all the measurements taken at energies $E_{cm}<2000\,keV\,$%
\thinspace \thinspace have been overestimated by at least $4\%$. A slight
screening enhancement has already appeared at $S\left( 1477\,keV\right) $
(see the relevant plot in Ref. \cite{harsr} ) which for most practical
purposes can be disregarded. Any attempt to lower the beam energy in order
to search for undetected resonances or improve the $S\left( 0\right) \,$%
value will cause considerable enhancement of the measurements, which can be
easily corrected through the screening enhancement factor given by Eq. $%
\left( \ref{psef}\right) $.

In absolute values, the screening errors which will be committed in future
CNO and rp astrophysical reactions are indeed important. However, their
actual importance can only be assessed when projected against the entire
background of experimental errors encountered in the laboratory. For
example, in the reaction $^{19}F\left( p,\alpha \gamma \right) ^{16}O\,$%
\thinspace the present model gives a screening error of order 5\%, while the
experimental uncertainty is 80\%. Likewise, the best available astrophysical
factor $S\left( 0\right) $ for the reaction $^{14}N\left( p,\gamma \right)
^{15}O$ carries a systematic error of 25\% when the respective screening
error given by the present model is of order 8\%. It seems therefore that
screening corrections for the astrophysical reactions in question can play a
significant role only if experimentalists manage to reduce other larger
experimental errors. This has been the case in the third experiment
discussed above (i.e. $^{86}Sr\left( p,\gamma \right) ^{87}Y)\,\,$where the
experimental uncertainty is of order $8\%\,\,$while the screening error is
of order $6.7\%.\,$

In conclusion, the present study gives a limit of errors in ignoring
electron screening. Its results indicate that, for most practical purposes,
screening corrections to past CNO and rp experiments can be disregarded.
Regarding future experiments, such as the ones conducted by the LUNA
collaboration, corrections to the CNO reactions will certainly be of
importance as they are closely related to the solar neutrino fluxes (via the 
$^{13}N\left( e^{+}\nu _{e}\right) ^{13}C$ and $^{15}O\left( e^{+},\nu
_{e}\right) ^{15}N\,$\thinspace reactions) and the rp process (via the
HotCNO cycle). However, it is up to experimentalists to judge the actual
importance of screening corrections by comparing them with all other
experimental errors encountered in the laboratory.

Finally, we should point out that:

a) While we have only investigated proton-induced astrophysical reactions,
the formalism of Ref . \cite{lioliossef} can be very easily applied to the
study of alpha-induced reactions where corrections are expected to be just
as important.

b) The importance of the effects of the ''solid state'' environment on
screening has been studied in a recent experiment\cite{rolfsddmetals} where
the usual puzzling result emerged again: The experimentally obtained
screening energy is larger than the one predicted by theoretical models. Due
to the importance of screening corrections and the future ambitious plans%
\cite{nupecc} in nuclear astrophysics experiments a more sophisticated study
is needed which should attempt to solve the relevant molecular few-body
problem.

{\bf ACKNOWLEDGMENTS}

The author would like to thank K.Langanke for useful comments and
discussion. This paper was inspired during the 2002 Symposium of the
Hellenic Nuclear Physics Society  where the author pointed out to the
collaborators of Ref. \cite{hars} that their experimental measurements have
been contaminated by screening effects.

\bigskip FIGURE CAPTIONS

Figure 1. The most effective energy of interaction $E_{0}\,$with respect to
plasma temperature for various proton-induced thermonuclear reactions $%
\left( p+_{Z}^{A}X\right) $with $Z_{1}=6,10,15,20,25,30,35,40\,$

Figure 2. The laboratory screening enhancement factor with respect to the
relative energy of interaction (center of mass) for various proton-induced
thermonuclear reactions $\left( p+_{Z}^{A}X\right) \,$with $%
Z_{1}=6,10,15,20,25,30,35,40\,$\thinspace

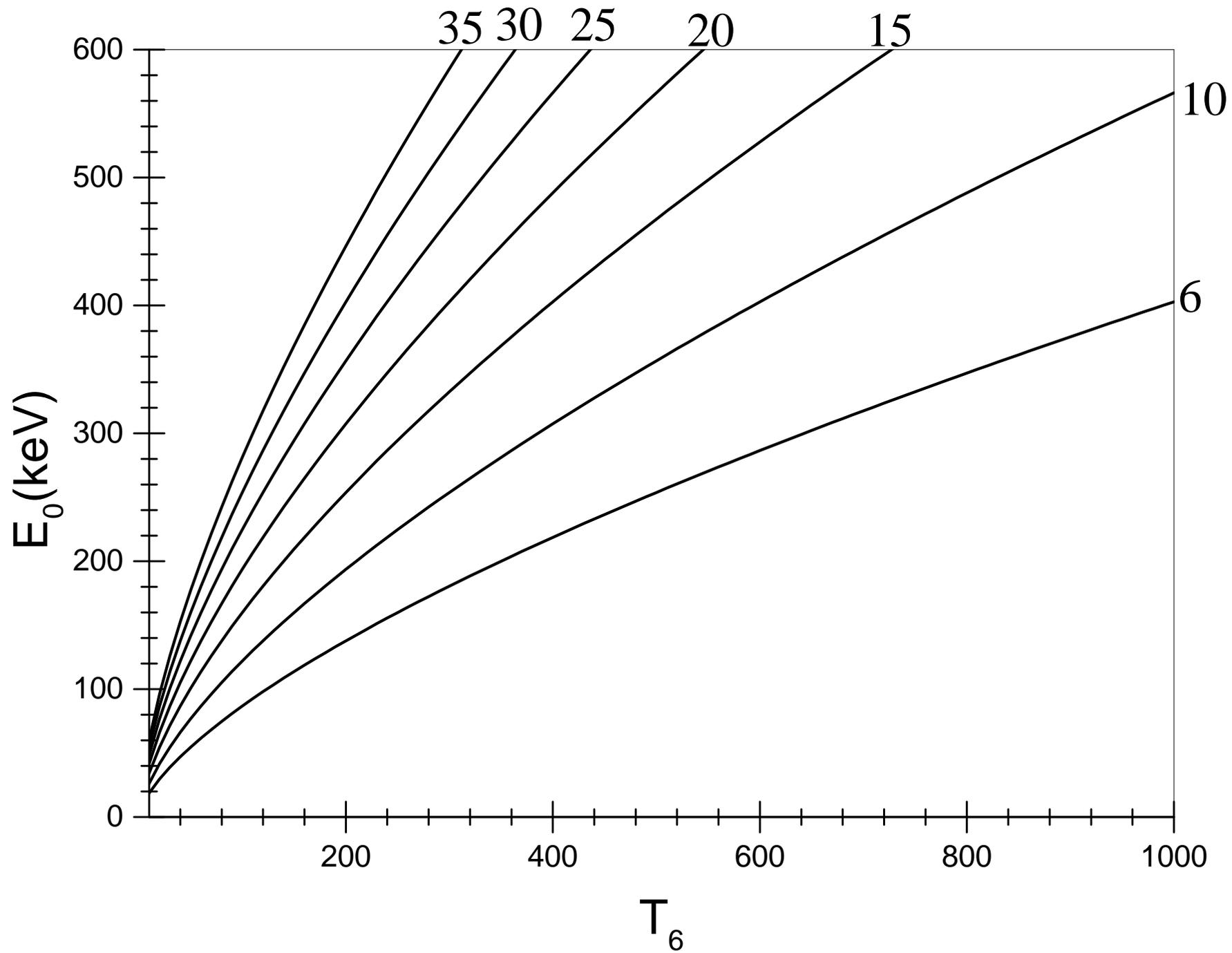
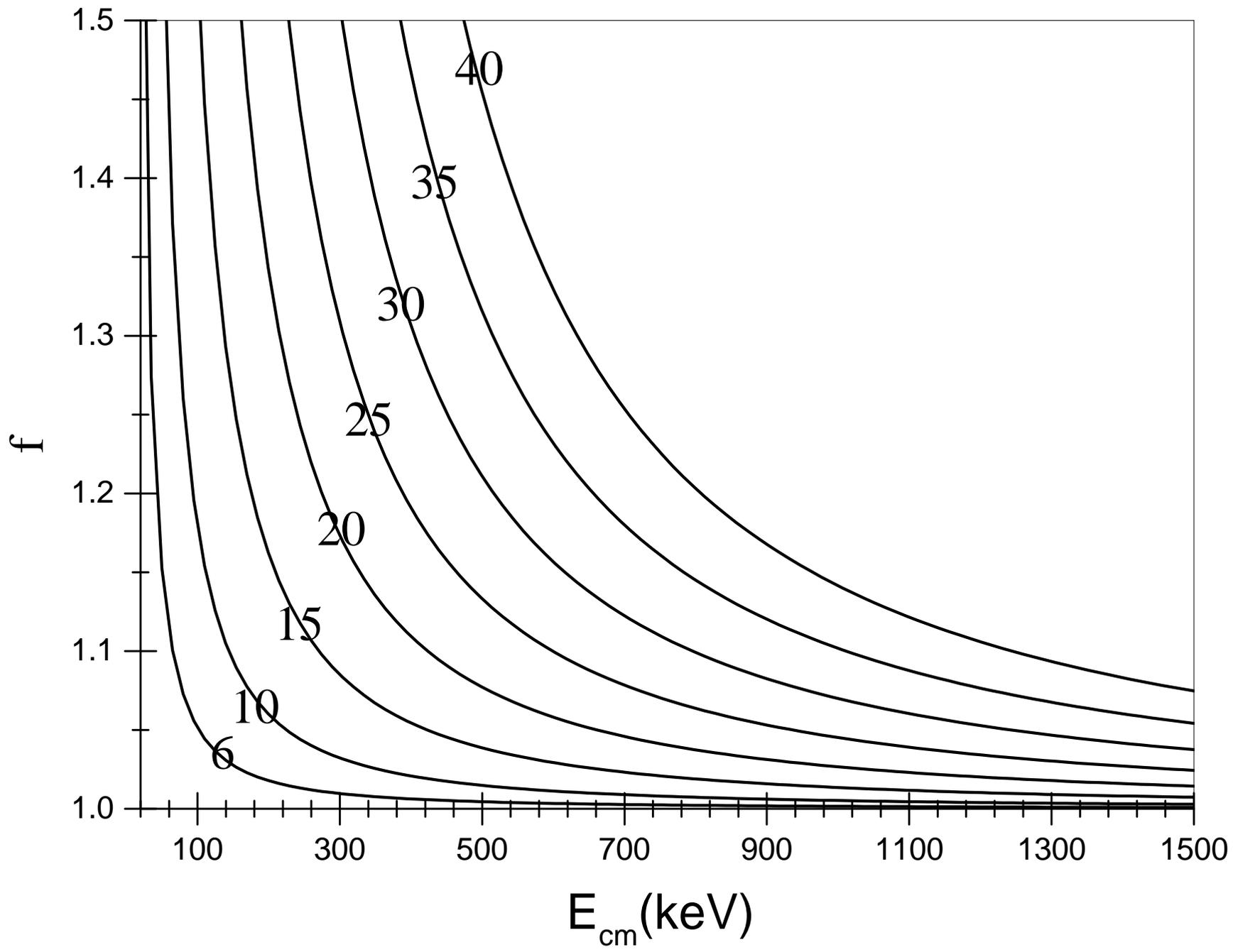
\end{document}